\documentclass[twocolumn,prl,showpacs]{revtex4}

\usepackage{graphicx,amssymb,amsmath}
\usepackage[applemac]{inputenc}

\begin{document}

\title{Are there waves in elastic wave turbulence ?}
\author{Nicolas Mordant}
\affiliation{Laboratoire de Physique Statistique, Ecole Normale Sup\'erieure \& CNRS, 24 rue Lhomond, 75231 Paris cedex 05, France}


\begin{abstract}
An thin elastic steel plate is excited with a vibrator and its local velocity displays a turbulent-like Fourier spectrum. This system is believed to develop elastic wave turbulence. We analyze here the motion of the plate with a two-point measurement in order to check, in our real system, a few hypotheses required for the Zakharov theory of weak turbulence to apply. We show that the motion of the plate is indeed a superposition of bending waves following the theoretical dispersion relation of the linear wave equation. The nonlinearities seem to efficiently break the coherence of the waves so that no modal structure is observed. Several hypotheses of the weak turbulence theory seem to be verified, but nevertheless the theoretical predictions for the wave spectrum are not verified experimentally.
\end{abstract}

\maketitle

Wave turbulence is a non equilibrium state in which dispersive wavetrains interact due to weak nonlinearities. The canonical example is  the case of gravity waves at the surface of the ocean~\cite{Hasselmann}. Many other systems belong to this class such as capillary waves, nonlinear optical waves, superfluids, Alfv\'en or atmospheric waves etc (see for example~\cite{Newell} for a short review). Contrary to the strongly nonlinear case of 3D hydrodynamic turbulence, a perturbative approach to weak turbulence was developed in the 60-70's. The idea of the theoretical development is, starting from the nonlinear wave equation, to develop a hierarchy of linear equations for the successive cumulants~\cite{Newell}. Using the hypothesis of weak nonlinearity, a temporal multiscale analysis can be developed which leads to a natural closure of this hierarchy. This is the main difference from hydrodynamics turbulence for which no such rigorous closure exists. From this closure, a kinetic equation can be derived describing the slow evolution of the Fourier components of the energy spectrum. The energy is slowly (compared to the natural frequency of the wave) exchanged among modes through the occurrence of resonances between $N$ waves. In the case of gravity waves, the resonance involves four waves, and three waves are resonant for capillary waves. 

Stationary solutions of these kinetic equations can be exhibited~\cite{Newell}. Those can be either equilibrium or out-of-equilibrium solutions. The former correspond to the Rayleigh-Jeans spectrum of equipartition of energy among the modes. The typical case where out-of-equilibrium solutions are expected is the following: the forcing of the system is applied at large scales and some additional dissipation mechanism such as diffusivity is operating at small scale to dissipate the energy. This is also a canonical situation of 3D hydrodynamics. The kinetic equation conserves energy (and some other invariant depending on the case under consideration).
% and in some cases like four waves resonance (e.g. surface gravity waves) another quantity can be conserved: the wave action. The former case
This leads to a direct energy cascade and a self-similar Kolmogorov-Zakharov spectrum with a constant energy flux across the scales. 
%The second case, when present, can lead to an inverse cascade with a constant particle flux.  

The weak turbulence theory assumes a certain number of hypotheses. Among these is the weak nonlinearity. Another is that the system is supposed to be large enough so that no effect of boundaries can be expected through the occurrence of cavity modes (the size of the system goes to infinity in the derivation). The goal of this article is to check some of these hypotheses in a real system: a thin elastic plate where bending waves are forced. Following recent work, this system is believed to be a good candidate for wave turbulence~\cite{During, Arezki}. We first describe our experimental setup and show estimations of the spectrum of the normal velocity of the plate. Finally, we investigate whether wave structure is indeed present in our system, estimate the dispersion relation of the observed waves, and relate our observations to some hypotheses of the weak turbulence theory. %\\[0.1cm]

\begin{figure}[!htb]
\centering
\includegraphics[height=5.5cm]{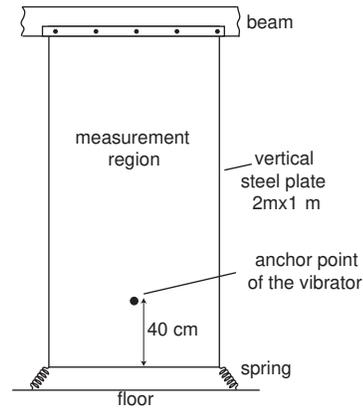}
\caption{Experimental setup: a 0.4~mm thick, $1\times2$~m$^2$ stainless steel plate is hanging under its own weight. It is bolted along the full length of one of its short sides on a beamer. Two soft springs hold the bottom corners loosely to prevent too large excursions of the plate. A vibrator is anchored on the plate at about 40~cm from the bottom and excites bending waves by moving normally to the plate. The normal velocity of the plate is recorded by a laser vibrometer  in a region slightly above the middle of the plate.}
\label{es}
\end{figure}
We report here a study of bending waves of a stainless steel plate, of size $2\times 1$~m$^2$ and $h=0.4$~mm thick, held vertically on a short side (Fig.~\ref{es}). A electromagnetic vibrator type V406 from LDS is fixed at a point located 40~cm from the plate bottom. It can move normally to the plate to excite bending waves. The typical forcing excitation is either sinusoidal (typically at 30~Hz) or a low frequency noise (of restricted bandwidth up to 15~Hz, for example). The forcing is recorded by a force probe of type NTC from FGP Sensors and an accelerometer 4393V from Br\"uel \& Kj{\ae}r. The motion of the plate is recorded by using a vibrometer type OFV-552 from Polytec. The vibrometer is equipped with a two-headed sensor that enables us to record the normal velocity of the plate at one point or normal velocity differences between two points. The measurement points are close to the center of the plate or in the upper half of the plate so that the motion of the plate is not directly affected by the vibrator.

This kind of thin elastic plate is known to develop bending waves following the wave equation~\cite{Landau,During}
\begin{eqnarray}
\frac{h^2E}{12(1-\sigma^2)}\Delta^2\zeta+\rho\partial_{tt}\zeta=\{\zeta,\chi\}\label{le}\\
\frac{1}{E}\Delta^2\chi=-\frac{1}{2}\{\zeta,\zeta\}\label{nle}
\end{eqnarray}
where $\zeta$ is the normal displacement and $E\simeq2\times10^{11}Pa$ is the Young modulus, $\sigma\simeq0.3$ is the Poisson ratio, and $\rho\simeq8000$~kg.m$^{-3}$ is the specific mass (values are typical for stainless steel). $\Delta$ is the Laplacian, and $\{a,b\}=\partial_{xx}a\,\,\partial_{yy}b+\partial_{yy}a\,\,\partial_{xx}b-2\partial_{xy}a\,\,\partial_{xy}b$. $\chi$ is the stress function and is related to the longitudinal deformation of the plate~\cite{Landau}.
The linear dispersion relation is 
\begin{equation}
\omega=\sqrt{\frac{Eh^2}{12(1-\sigma^2)\rho}}k^2\, .
\label{rd}
\end{equation}
 For large amplitudes, the wave equation exhibits cubic nonlinearities. This nonlinear equation is believed to lead to wave turbulence \cite{During} with the following predicted wide-band spectrum for the displacement $\zeta$:
\begin{equation}
E_\zeta(k)=C\frac{P^{1/3}}{\left[12(1-\sigma^2)\right]^{1/6}}\frac{\ln^{1/3}(k^\star/k)}{\sqrt{E/\rho}k^3}\, ,
\label{during}
\end{equation}
where $P$ is the energy flux per unit mass (here we use the definition of Connaughton {\it et al.}~\cite{Conn}), $C$ a dimensionless number, and $k^\star$ is an {\it ad hoc} cutoff wave number. Using (\ref{rd}) to change variables from $k$ to $\omega$, the theoretical prediction for the velocity power spectrum is then
\begin{equation}
E_v(\omega)=C^{\prime}\frac{hP^{1/3}}{\left[12(1-\sigma^2)\right]^{2/3}}\ln^{1/3}(\omega^\star/\omega)\, ,
\label{tsv}
\end{equation}
where $\omega^\star=\omega(k^\star)$ and $C^{\prime}$ is a number.

Following Connaughton {\it et al.}~\cite{Conn}, another prediction, based on dimensional analysis, can be derived: Starting from the linear dispersion relation of the form $\omega=\lambda k^\alpha$, using the order of the nonlinearity to get the scaling in $P$ and assuming power laws, one predicts a velocity power spectrum
\begin{equation}
E_v(\omega)\sim P^{1/3}\left(\frac{Eh^2}{12(1-\sigma^2)\rho}\right)^{1/4}\omega^{-1/2}
\end{equation}
for four-wave interactions (here $\lambda=\sqrt{\frac{Eh^2}{12(1-\sigma^2)\rho}}$).

In order to get a more general prediction based on dimensional analysis, one should note that the Young modulus appears in the linear part of the wave equation (\ref{le}) combined with $h$ and $\rho$  but also alone in the nonlinear part (\ref{nle}). This is different from the case of surface gravity waves for example, for which no additional dimensional parameters appear in the nonlinear part, due to a different physical nature of the nonlinearity.
Here, dimensional analysis should use $E$, alone and not combined in $\lambda$ with $h$ and $\rho$. In this way, with the parameters $P$, $E$, $\rho$ and $h$ (discarding $\sigma$ as without dimension), one predicts that the velocity power spectrum should behave like:
\begin{equation}
E_v(\omega)= h\sqrt{\frac{E}{\rho}}\,\,g\left(\frac{P^{1/3}}{h\omega},P\left(\frac{\rho}{E}\right)^{3/2}\right)\, ,
\label{sca}
\end{equation}
where $g$ is an unknown function. The theoretical prediction (\ref{tsv}) follows the dimensional prediction (\ref{sca}) with $g(a,b)\propto (b\ln a)^{1/3}$. One would then expect the cutoff frequency to follow 
\begin{equation}
\omega^\star \propto P^{1/3}/h.
\label{co}
\end{equation} 
In \cite{During}, this high frequency cutoff $\omega^\star$ had to be introduced in order to ensure that a non-zero energy flux exists in the out-of-equilibrium case. In the dimensional analysis, it appears naturally from the nonlinear conservative wave equation. Note that in the case of surface gravity waves, one has to invoke dissipation (or the crossover to capillary waves) to introduce a high-frequency cutoff in the wave spectrum.

In our case, $P$ can be estimated experimentally. The mechanical power $\mathcal P$ transferred to the plate by the vibrator can be estimated by $\mathcal P=\langle \mathcal F U\rangle$, where $\mathcal F$ is the force applied to the plate and $U$ the velocity of the plate at the vibrator position. For a given plate, $P$ is proportional to $\mathcal P$.

\begin{figure}[!htb]
\centering
\includegraphics[width=8cm]{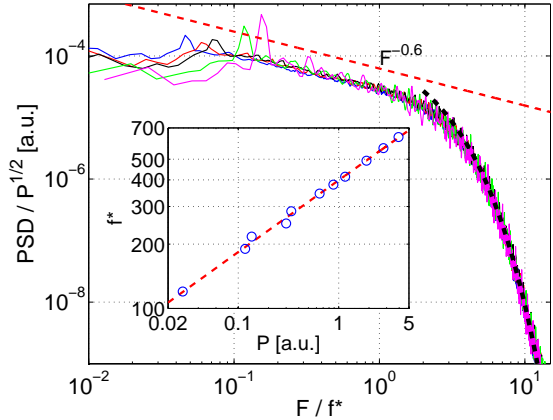}
\caption{Power spectrum density of the plate normal velocity. The frequency is normalized by a cutoff frequency $f^\star$ estimated from the high frequency exponential decay region. The spectrum is normalized by $\mathcal P^{1/2}$. The forcing is sinusoidal at 30~Hz and 5 curves  are displayed for $\mathcal P=3.94$, $1.94$, $1.16$, $0.30$ and $0.117$~W (from left to right). The dashed line is a $F^{-0.6}$ decay as an eye guide. The thick dashed line is an pure exponential decay. Insert: Evolution of the cutoff frequency $f^\star$ with $\mathcal P$. The dashed line is a $\mathcal P^{1/3}$ fit. }
\label{ps}
\end{figure}
Figure~\ref{ps} displays the velocity power spectrum measured at a point near the center of the plate. At the highest frequencies, the spectrum decays exponentially (thick dashed line). A cutoff frequency can then be estimated by an exponential fit of the spectrum in this high-frequency region. As seen in the inset, $f^\star$ varies as $P^{1/3}$ in agreement with (\ref{co}). This suggests that the observed cutoff may not be due to any dissipative process but rather to the nonlinear dynamics of the waves (although some dissipative process eventually absorbs the energy flux).

In the main figure, the frequency has been rescaled by $f^\star$ (i.e., $P^{1/3}$) and the amplitude by $P^{1/2}$ showing a good collapse as suggested previously by Boudaoud {\it et al.}~\cite{Arezki}.  The full spectrum, including the high frequency cutoff region, seems to follow a master curve. An inertial region could be extracted at frequencies just above the excitation frequency for which a power law decay can be fitted. The best fit exponent is close to -0.6. Altogether this leads to a scaling $E_v(\omega)\propto P^{0.7}\omega^{-0.6}$ which is quite far from the theoretical prediction $E_v(\omega)\propto P^{1/3}\omega^{0}$ of D\"uring {\it et al.}~\cite{During}. As discussed by \cite{Arezki}, the scaling exponent  in $P$ is closer to that of a three-wave interaction. Note that the $\omega^{-0.6}$ scaling is not far from the dimensional prediction by Connaughton {\it et al.}~\cite{Conn} for a four-wave interaction. For three waves, their argument would lead to $\omega^{-3/4}$, so the actual observed exponent lies in between the two predictions. Boudaoud {\it et al.} suggest that this discrepancy of the scaling in $P$ could come from the fact that the plate is not ideal, which is certainly true here: in particular it is not perfectly flat at rest. This can induce quadratic nonlinearities that would cause three-wave interactions~\cite{Arezki}. They also suggest that some singular structures like d-cones or ridges may be dominant in the dynamics of the plate.

To tackle the question of the persistence of waves in this nonlinear system, we used measurements of the velocity differences between two points separated spatially in the plate, namely $\delta v(x)=v(x,t)-v(0,t)$, where $x$ is the coordinate along the axis of the rectilinear displacement of one arm of the two-headed sensor of the vibrometer (the other arm being at position $0$). Among the hypotheses required for the derivation of the Kolmogorov-Zakharov spectrum of weak turbulence, one assumes that the wave structure is conserved, that the linear dispersion relation remains valid, and that the size of the system is large enough so that no modal structure due to the boundaries is retained. In this spirit, one can assume that the motion of the plate is governed by a superposition of pure sine waves that remain coherent over the length of the measurement so that one can write for the Fourier component of $\delta v$ at frequency $\omega$:
\begin{equation}
\delta v(x,\omega)=e^{-i\omega t}\int A(\theta,\omega)(e^{ik\cos\theta x}-1)kd\theta
\end{equation}
where $A(\theta,\omega)$ is the amplitude of the wave traveling with an angle $\theta$ to the axis of the measurement.
The frequency spectrum is then 
\begin{equation}
\langle |\delta v(x,\omega)|^2\rangle=\langle\left|\int A(\theta,\omega)(e^{ik\cos\theta x}-1)kd\theta\right|^2\rangle
\label{eq1}
\end{equation}
%qui peut s'écrire
%\begin{equation}
%\langle |v(x,\omega)-v(0,\omega)|^2\rangle=k^2\iint \langle A(\theta,\omega)A^\star(\theta^\prime,\omega)\rangle (e^{i(k\cos\theta x)}-1)(e^{-i(k\cos\theta^\prime x)}-1)d\theta d\theta^\prime
%\end{equation}
%\begin{equation}
%\langle A(\theta,\omega)A^\star(\theta^\prime,\omega)\rangle=\langle|A(\omega)|^2\rangle\delta(\theta-\theta^\prime).
%\end{equation}

If one assumes that correlations among waves require an interaction of more than just two waves (as would be expected for the four-wave interaction suggested in this case) then one can write, with the additional assumption of isotropy: 
\begin{equation}
\langle A(\theta,\omega)A^\star(\theta^\prime,\omega)\rangle=\langle|A(\omega)|^2\rangle\delta(\theta-\theta^\prime)\, .
\label{cor}
\end{equation}

Using (\ref{cor}) in (\ref{eq1}), one obtains:
\begin{equation}
\langle |\delta v(x,\omega)|^2\rangle=4\pi k^2\langle|A(\omega)|^2\rangle (1-J_0(kx))\, .
\end{equation}
The prefactor of the $J_0$ Bessel function is simply twice the spectrum of the single point velocity $\langle|v(\omega)|^2\rangle$ so that 
\begin{equation}
E_{\delta v}(x,\omega)=2E_v(\omega)(1-J_0(kx))
\label{fit}
\end{equation}
This result is well known in the study of diffuse fields in acoustics and the $J_0$ function that arises here is related to the Green function of the wave in free space~\cite{Weaver}.

\begin{figure}[!htb]
\centering
\includegraphics[width=8cm]{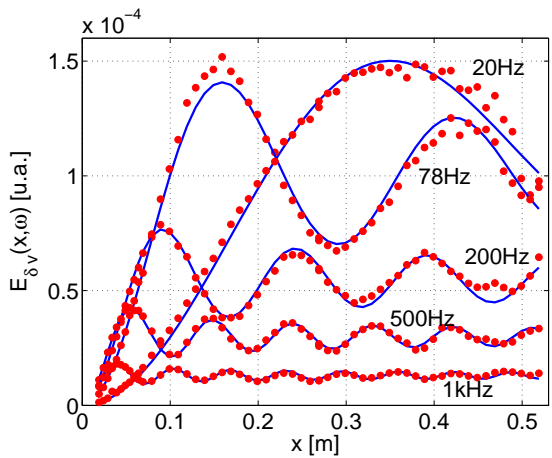}
\caption{Mixed Fourier spectrum $E_{\delta v}(x,\omega)$ of the velocity differences $\delta v$ (dots) compared to a fit by a $J_0$ Bessel function (continous line) for various values of the frequency. The forcing is sinusoidal at 30~Hz.}
\label{tf}
\end{figure}

Here, we measure sequentially $\delta v(x,t)$ for values of $x$ ranging from 1.8~cm to 50~cm. The temporal spectrum of each time series is estimated, and all spectra are gathered to obtain $E_{\delta v}(\omega,x)$. Examples of these spectra are shown in Fig.~\ref{tf} for various values of the frequency. A $J_0$ Bessel function is then fitted to the data. One can see that the fit is very good for all frequencies. At high frequencies (above 2~kHz typically), the spatial oscillation is not adequately resolved by our measurement spatial spacing of 1~cm and sufficient statistical convergence may not be attained, as the amplitudes decay quite fast for these frequencies. 

\begin{figure}[!htb]
\centering
\includegraphics[width=8cm]{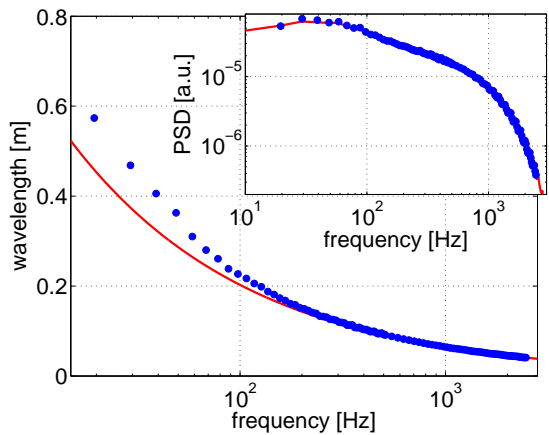}
\caption{Evolution of the wavelength as a function of the frequency. Dots: experimental data, continuous line: theoretical value from (\ref{rd}). Insert: estimation of the spectrum of the normal velocity $E_v(\omega)$ from a direct measurement (average over the same positions as the two point measurements) -- solid line -- and extracted from the fit parameter of the mixed spectrum $E_{\delta v}(x,\omega)$ -- dots. Same dataset as in fig.~\ref{tf}.}
\label{dr}
\end{figure}

The spatial scaling factor is shown as a function of the frequency and compared to the theoretical wavelength in Fig.~\ref{dr}. The fitted amplitude is also displayed and compared to an estimate of the spectrum measured at a single point.
The estimated dispersion relation is very close to the theoretical one (\ref{rd}). A systematic deviation of the experimental data is observed at low frequencies. For frequencies over 100~Hz, the mismatch is below 10$\%$ and over 500~Hz it is negligible compared to all uncertainties. The mismatch at low frequency can be due to systematic errors, as very few oscillations of the mixed spectrum are observed, so that the fitting of the $J_0$ function may cause systematic errors.  At low frequency, the wavelength is also comparable with the size of the plate, so that the presence of the springs and the fact that the plate is held on the vibrator may influence strongly the dynamics of the waves. Note that the estimations performed for several amplitudes of forcing show no visible variation of the wavelength estimation (not shown). This means that no influence of the nonlinearity is observed on this quantity, which is one of the requirements for the weak turbulence theory to apply.

The velocity power spectrum estimated from the fit of the mixed spectrum $E_{\delta v}(x,\omega)$ by the $J_0$ Bessel function  (\ref{fit}) is compared in the insert of Figure~\ref{dr} with a direct estimate of $E_v(\omega)$. The latter is obtained by averaging the spectrum $E_v(\omega)$ over the same positions $x$ used to estimate $E_{\delta v}(x,\omega)$. One can see that the two estimations are coincident at all frequencies. 

From all these observations, one can conclude that the motion of the plate is indeed caused by a superposition of waves. The wavelength variation as a function of the frequency follows quite closely the theoretical dispersion relation and no strong nonlinear effect is observed at this level. The spectrum of velocity differences follows very closely the behavior expected for isotropic waves in an infinite medium and no mode structure is evidenced. All these facts support the hypotheses required for the theory of weak turbulence to apply. Nevertheless a mismatch is observed between the measured and theoretical spectrum. The scaling of the spectrum with the injected power is in strong disagreement with the prediction. The analytical functional form of the spectrum is also not observed. Our two-point measurement support the fact that the dynamics of the plate is due to isotropic waves and not to more complex dominant structures like ridges or d-cones as suggested in~\cite{During,Arezki}. The observation of waves also supports the usual change of variable from wavenumber  $k$ to frequency $\omega$ through the dispersion relation. This is the equivalent for wave turbulence of the Taylor hypothesis of hydrodynamic turbulence that relates space scales to time scales.

\begin{acknowledgments}
This work is supported by the ANR ``TURBONDE" contract. The author would like to thank S. Fauve, E. Falcon, C. Falc\'on, B. Gallet, S. Auma\^{\i}tre, C. Laroche, S. Rica, G. D\"uring, and C. Prada for numerous discussions related to this work. 
\end{acknowledgments}

 \end{document}